\documentclass[journal]{IEEEtran}

\usepackage[latin1]{inputenc}
\usepackage{graphicx,textcomp}
\usepackage[cmex10]{amsmath}
\usepackage{amssymb,subfigure}
\usepackage{hhline,color,siunitx}
\usepackage{blindtext}
\usepackage{esint}
\usepackage{url}

\usepackage[noadjust]{cite}

\newlength\figureheight
\newlength\figurewidth
\begin{document}

  \renewcommand{\arraystretch}{1.3}

%
\title{Electromagnetic Modeling of Superconductors with Commercial Software: Possibilities with Two Vector Potential-Based Formulations}

\author{Francesco Grilli, Enric Pardo,~\IEEEmembership{Senior Member,~IEEE}, Antonio Morandi,~\IEEEmembership{Senior Member,~IEEE}, \\ V\'ictor M. R. Zerme{\~n}o, Roberto Brambilla, Tara Benkel, Nicol\`o Riva
\thanks{Document written on \today.}
\thanks{F.~Grilli and T.~Benkel are with the Karlsruhe Institute of Technology, Germany.
A.~Morandi is with the University of Bologna, Italy. E. Pardo is with the Slovak Academy of Science, Bratislava, Slovakia. V. M. R. Zerme{\~n}o is with NKT, Cologne, Germany. R.~Brambilla was with RSE, Milano, Italy. N. Riva is with the \'Ecole Polytechnique F\'ed\'erale de Lausanne, Switzerland.}
\thanks{Corresponding author's email: francesco.grilli@kit.edu.}
}



\maketitle

\begin{abstract}
In recent years, the $H$ formulation of Maxwell's equation has become the de facto standard for simulating the time-dependent electromagnetic behavior of superconducting applications with commercial software. However, there are  cases where other formulations are desirable, for example for modeling superconducting turns in electrical machines or situations where the superconductor is better described by the critical state than by a power-law resistivity.
In order to accurately and efficiently handle those situations, here we consider two published approaches based on the magnetic vector potential: the $T$-$A$ formulation of Maxwell's equations (with power-law resistivity) and Campbell's implementation  of the critical state model. In this contribution, we extend the $T$-$A$ formulation  to thick conductors so that large coils with different coupling scenarios between the turns can be considered. We also revise Campbell's model and discuss it in terms of its ability to calculate AC losses: in particular, we investigate the dependence of the calculated AC losses on the frequency of the AC excitation and the possibility of using quick one-step (instead of full cycle) simulations to calculate the AC losses.
\end{abstract}


\IEEEpeerreviewmaketitle


\section{Introduction}
\IEEEPARstart{N}{umerical} models have become popular tools for understanding the behavior of superconductors and for designing  applications. Among the models used for investigating the electromagnetic behavior of superconductors, the finite-element method (FEM) based on the $H$ formulation of Maxwell's equations combined with the power-law model of the superconductor is by far the most widely adopted approach, used by tens of research group around the world~\cite{Shen:SST20}.
The reason of such popularity mainly resides in the easiness of implementation in the  FEM program Comsol Multiphysics~\cite{Hong:SST06,Brambilla:SST07}, although implementations in other commercial software  packages  like FlexPDE~\cite{Lorin:TAS13} and Matlab~\cite{Lahtinen:SST12}, open-source environments like GetDP~\cite{Makong:TMAG18a},   and   home-made   FEM   codes   like   Daryl Maxwell~\cite{Escamez:TMAG16} also exist.

In this contribution, we discuss two approaches based on the magnetic vector potential, which -- for different reasons -- can be considered as an alternative to the $H$ formulation for some application contexts.

The $T$-$A$ formulation, proposed by Zhang et al. in~\cite{Zhang:SST17}, is becoming a popular tool for solving electromagnetic problems involving HTS coated conductors which can be treated as infinitely thin objects~\cite{Liang:JAP17,Wang:SST19,Yan:TAS19,BerrospeJuarez:SST19}. This model too uses the power-law as constitutive relation of the superconductor. Here, we extend the formulation to thick superconductors: not only does this allow simulating other types of superconducting tapes (like Bi-2223 or $\rm MgB_2$ flat rectangular tapes), but -- perhaps more importantly -- it also allows simulating stacks of electromagnetically coupled coated conductors, which are often used in high-current HTS cables~\cite{Takayasu:SST12,Celentano:TAS14,Uglietti:TAS14}. In stacks of coated conductors, the superconducting layers of the various tapes are electromagnetically coupled and the whole stack can be assimilated to a thick superconductor. One advantage of this formulation is that it can be directly used to simulate HTS in electrical machines, if those are modeled with a formulation based on the magnetic vector potential $A$~\cite{Benkel:arXiv19}. 

Numerical formulations using the vector potential $A$ can be also combined with different (from the power law) constitutive models of the superconductor. In this paper  the quasi critical state model (QCSM) proposed by A. M. Campbell in~\cite{Campbell:SST07} is used for obtaining a fast solution in terms of vector potential $A$ by solving a backward sequence of non-linear magnetostatic problems.
Here we show that, due to the smoothness with which the current density switches between $+J_{\rm c}$ and $-J_{\rm c}$, the model is not fully rate-independent (hence the proposed name). We also show that, at least in certain cases, a one-step calculation of the field distribution corresponding to the peak of the AC excitation can be used to rapidly calculate the cyclic AC losses of individual superconductor tapes.

\section{Numerical Models}
All the models considered in this article are 2D models simulating the cross section of superconductors in the $xy$ plane, with the current flowing in the $z$ direction (Fig.~\ref{fig:TA_1D_2D}). The superconductors are considered to be infinitely long in the $z$ direction.
\subsection{T-A Formulation with Power-Law Model}\label{sec:T-A}
{The $T$-$A$ formulation was proposed as a means to tackle the computational challenge of simulating HTS coated conductors, which are characterized by a superconducting layer with very large width-to-thickness ratio. The HTS tapes are modeled as 1D objects and the current vector potential $\mathbf{T}$ is used as state variable in Faraday's equation. The magnetic field in all the simulated domains (including non-superconducting regions) is calculated with the magnetic vector potential  $\mathbf{A}$ formulation. The two formulations are coupled, so that the electromagnetic interaction between multiple tapes can be calculated.
In  the $T$-$A$ formulation, the current $I$ flowing in a conductor of cross section $S$ is given by
\begin{equation}\label{eq:current}
I=\iint_S {\mathbf J}\,{\rm d}S=\iint_S \nabla \times {\mathbf T}\,{\rm d}S=\oint_L{\mathbf T} \,{\rm d}l,
\end{equation}
where $L$ represents the boundary edges of the cross section $S$.
Then,  the main difference between thin and thick superconductors is in the way to impose such condition.
In thin conductors, the current is imposed by setting appropriate (0D) boundary conditions at the extremities of each tape, as explained in~\cite{Zhang:SST17}.
In thick conductors of rectangular cross section, different sets of conditions can be set for the two components of the current vector potential, $T_x$ and $T_y$. Four examples are represented in Fig.~\ref{fig:TA_1D_2D}. One can easily verify that they are all consistent with equation~\eqref{eq:current} for imposing a current $I$ of desired amplitude. This way of implementing equation~\eqref{eq:current} takes advantage of the rectangular geometry: for example the projection of $T_y$ onto the top and bottom boundaries of the rectangle (which are parallel to the $x$ direction) is automatically zero. For more general shapes of the cross section, the implementation of equation~\eqref{eq:current} is less straightforward, and the details of the implementation in COMSOL Multiphysics are given in the appendix.
\begin{figure}[t!]
\centering
\includegraphics[width=\columnwidth]{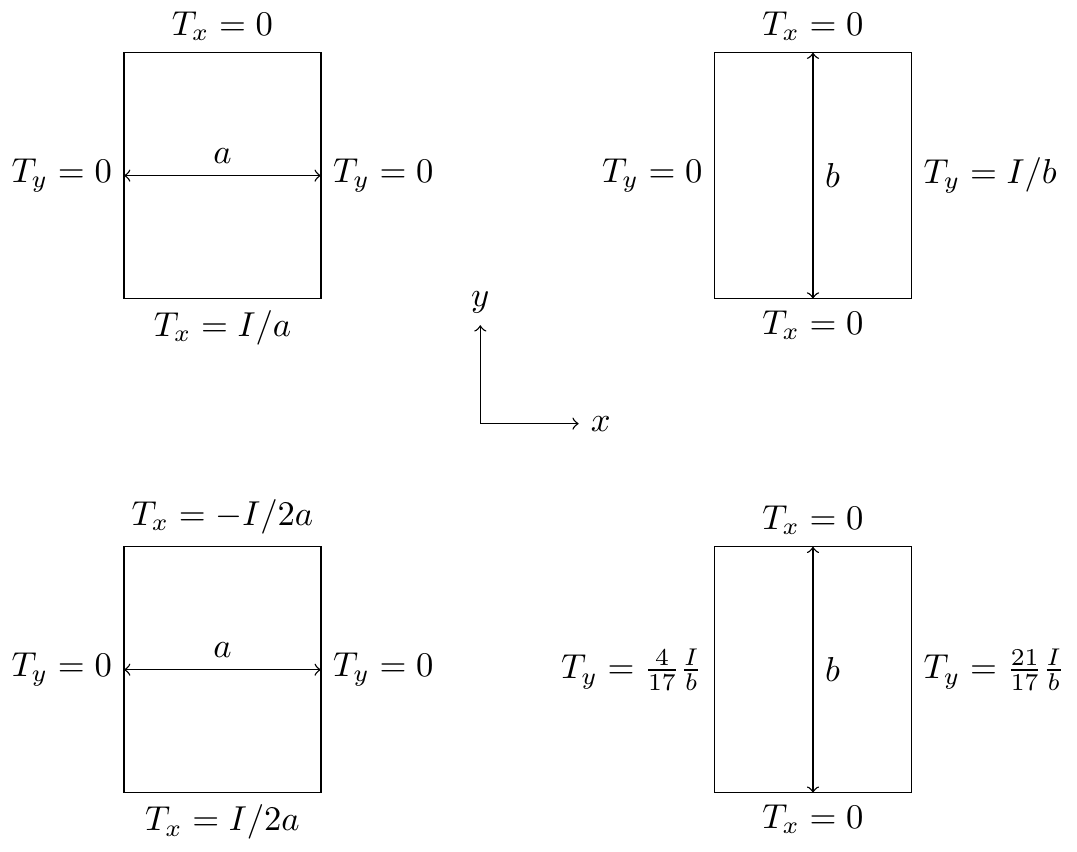}
\caption{\label{fig:TA_1D_2D}Four possible boundary conditions for the current vector potential $\bf T$ for imposing a transport current $I$ in a rectangular conductor of width $a$ and thickness $b$, according to equation~\eqref{eq:current}.}
\end{figure}
The $T$-$A$ formulation is implemented in COMSOL Multiphysics, by using the {\it PDE-Coefficient form} module for the $T$ part and the {\it Magnetic Fields} module for the $A$ part, respectively. The $T$ and $A$ parts use Lagrange first and second order elements, respectively. A discussion on the use of elements of different order can be found in the appendix of~\cite{BerrospeJuarez:SST19}.

The superconductor is modeled as a material with power-law resistivity
\begin{equation}\label{eq:pl}
\rho(J)=\frac{E_{\rm c}}{J_c}\left ( \frac{|\mathbf J|}{J_{\rm c}}\right)^{n-1},
\end{equation}
where $E_{\rm c}$ is the critical electric field, $J_{\rm c}$ is the critical current density and $n$ the power-law exponent defining the steepness of the $E$-$J$ curve.}
 
\subsection{$A$-Formulation with Quasi Critical State Model}\label{sec:QCSM}
The quasi critical state model (QCSM) solves the equation
\begin{equation}\label{eq:static_eq}
\nabla^2 {\mathbf A}=-\mu_0 {\mathbf J}
\end{equation}
where  $J$ takes values approximating the transition between $+J_{\rm c}$ and $-J_{\rm c}$ or zero. 
In the version discussed here (implemented in COMSOL Multiphysics), we model this transition as 
\begin{equation}\label{eq:EJ}
J=J_{\rm c} {\rm erf}\left ( \frac{E}{E_0}\right),
\end{equation}
$\rm erf$ is the error function~\cite{Glaisher:LEDPMJS1871}  and $E_0$ is a parameter defining the steepness of the switch between $+J_{\rm c}$ and $-J_{\rm c}$ (or from 0 to $\pm J_{\rm c}$ for virgin points).
In this work, we used $E_0=\SI{1e-6}{\volt\per\meter}$. 
Instead of $\rm erf$, other functions based on exponentials~\cite{Campbell:SST07} or hyperbolic tangent~\cite{Gomory:TAS12}  can be used to smooth the transition. 
The model is defined here as a {\it quasi} critical state model because it uses a smooth $E$-$J$ characteristic -- Eq.~\eqref{eq:EJ} -- , which results in a different behavior than that of a `pure' critical state model (prescribing sharp shift to $\pm J_{\rm c}$ produced by a non zero electric field regardless of its magnitude), as it will be illustrated later.

In addition, since we consider 2D problems, the magnetic vector potential has only one component (along $z$ in Fig.~\ref{fig:TA_1D_2D}), and from now on it will be treated as a scalar. The electric field driving the current originates from the time-variation of the magnetic vector potential $A$ plus a voltage gradient~\cite{Campbell:SST07,Gomory:TAS12}. In the present work, we consider a single isolated tape and the voltage gradient term can be mostly ignored although it will still influence the boundary conditions. In this way, we are  assuming Weyl's gauge, were ${\bf A}={\bf A}_c+\nabla \int {\rm d}t\phi$, being ${\bf A}_c$ and $\phi$ the vector potential in Coulomb's gauge and the electrostatic scalar potential, respectively. With this gauge
\begin{equation}\label{eq:E_deltaA_deltat}
E=-\frac{\partial A}{\partial t}\approx -\frac{A_{t+\Delta t}-A_t}{\Delta t}.
\end{equation}
By substituting~\eqref{eq:E_deltaA_deltat} and~\eqref{eq:EJ} in~\eqref{eq:static_eq}, we finally obtain 
\begin{equation}
\nabla^2 A (t+\Delta t)=-\mu_0 J_{\rm c} {\rm erf} \left ( -\frac{A_{t+\Delta t}-A_t}{E_{\rm 0} \Delta t}\right ).
\end{equation}
This equation, which corresponds to the backward Euler solution of non-linear and time-dependent problem, allows solving the time evolution of $A$  by simulating a series of static problems (one for each time step).
An external magnetic field or a transport current is imposed by setting the appropriate conditions for the magnetic vector potential on the boundary of the air domain surrounding the superconductor. For example, in 2D cartesian coordinates, a boundary condition
\begin{equation}\label{eq:boundary_condition}
A=B_0(-x \cos\theta+y \sin \theta)\sin(\omega t)
\end{equation}
generates a magnetic field of amplitude $B_0$,                                 angle $\theta$ with respect to the $y$ axis, and sinusoidal time dependence. A boundary condition $A=A_0\sin(\omega t)$, where $A_0$ is a constant, generates a sinusoidal transport current in the superconductor. The value of the current can be calculated {\it a posteriori} in the post-processing, by integrating $J$ over the superconductor's cross section at the peak of the current.

In the article where the quasi critical state model was originally proposed~\cite{Campbell:SST07}, it was mentioned that, in the case of an AC excitation, the superconductor's  cyclic losses could be simply computed by knowing the current density $J_{\rm p}$ and the magnetic vector potential $A_{\rm p}$ at the peak of the excitation as
\begin{equation}\label{eq:peak_losses}
Q=-4\int\limits_{\Omega} J_{\rm p} A_{\rm p} {\rm d}\Omega,
\end{equation}
where $\Omega$ is the superconductor's domain.

This expression was also mentioned in~\cite{Claassen:APL06, Clem:SST07,Pardo:SST07}.
However, as pointed out in section 2.5 of~\cite{Pardo:SST07} and in section II.C.2 of~\cite{Grilli:TAS14a}, its applicability for computing the cyclic AC losses is limited to certain conditions. 
First, we use Weyl's gauge, where ${\bf A}={\bf A}_c+\nabla\int{\rm d}t\phi$. Using this relation, we can see that equation (8) is equivalent to (20) in~\cite{Grilli:TAS14a} for Coulomb's (or any other) gauge. Second, this equation assumes that at each half-cycle the current density fronts penetrate monotonically from all external surfaces inwards, and hence the region with $J=+J_c$ grows towards that of $J=-J_c$, and vice versa. It is also necessary that at the initial stage the current fronts penetrate only towards the current-free kernel, where A vanishes in Weyl's gauge. This gauge is satisfied because, first, $J=0$ causes  $E=0$, and hence $\partial_tA$=0 and, second, $A=0$ initially and $\partial_tA=0$ follows from the beginning of the curve, so that $A$ remains null.
As the field increases from zero to the peak, the current density of the points of the superconductor for which $J$ is equal to $+J_{\rm c}$ or $-J_{\rm c}$ never changes (until when the field or the current is reversed). 
Examples of scenarios when this is not the case are combinations of simultaneous alternating transport current and magnetic field~\cite{Pardo:SST07} or the magnetization of a superconductor of elliptical cross section with inclined field~\cite{Wolsky:SST08}.
In the latter case, the problem stems from the fact that, while the increase of the field from zero to the peak is monotonic, the evolution of the current density in the superconductor is not: in other words, due to the deformation of the field lines inside the superconductor as the field is increased, some points inside the superconductor may switch between $+J_{\rm c}$ and $-J_{\rm c}$ (or vice versa) during the field ramp from zero to the peak value.
In Section~\ref{sec:QCSM_results}, we will verify this and try to assess the magnitude of the error committed by the one-step calculation and equation~(\ref{eq:peak_losses}) for calculating the cyclic AC losses.

\subsection{Other Models Used for Comparison}
The $T$-$A$ formulation with power-law and the $A$ formulation with the quasi critical state model are validated with a comparison with other models: the Minimum Electro-Magnetic Entropy Production (MEMEP) model and a `pure' critical state model, respectively. This subsection quickly summarizes these two models.

The MEMEP model uses the current density as state variable, avoiding meshing the air. Differently from integral methods, it solves $\mathbf J$ by minimizing a certain functional~\cite{Pardo:SST15,Pardo:JCP17}. This method can take any $E(J)$ relation into account, including the multi-valued relation of the CSM~\cite{Pardo:JCP17}. However, in this article we use the power-law $E(J)$ relation defined by the resistivity in~\eqref{eq:pl}.

As for the `pure' CSM, a sharp shift to $\pm J_{\rm c}$ is produced by a non-zero electric field regardless of its magnitude. This means that only the sign of the electric field rather than its magnitude determines the electrodynamics of the system. For calculating the numerical solution arising form the CSM assumption,  we follow the approach developed in~\cite{Morandi:SSt15} based on the $A$ formulation. A matrix equation involving the current density of the elements as state variable is introduced for the discretized problem. An iterative procedure is applied for solving this matrix equation subject to the constraint $|{\mathbf J}|=0, J_{\rm c}$ at any point of the superconductor. We emphasize that in order to obey to the pure critical state model we exactly impose the constraint  $|{\mathbf J}|=\{0, J_{\rm c}\}$ and do not replace it with $|{\mathbf J}| \le J_{\rm c}$. We also emphasize that~\eqref{eq:E_deltaA_deltat} represents a good mathematical representation of this statement as far as the problem is dominated by a sufficiently high electric field, arising from an intense time derivative of excitation (related to boundary condition~\eqref{eq:boundary_condition}) due to high frequency and/or high magnitude. This means that in this operating conditions the results of the QCSM and the pure CSM coincide. However, in the low electric field regime (which can occur when a low frequency or a small ripple current is considered), the two models differ, as it will become clear from the results shown in section~\ref{sec:QCSM_results}.

\section{Results}
\subsection{$T$-$A$ Formulation: Validation and Application to Electrical Machines}
\begin{figure}[t!]
\centering
\includegraphics[width=8 cm]{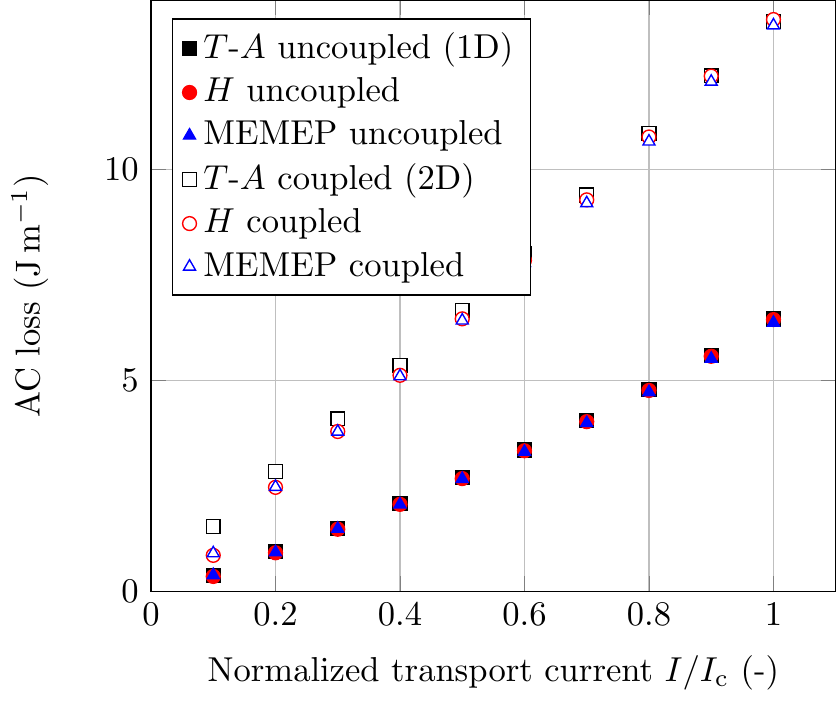}
\caption{\label{fig:coils_alone_plot}{AC losses as a function of the transport current (normalized to $I_{\rm c}$) of the racetrack coil considered in~\cite{Pardo:TAS19,Benkel:arXiv19}, when the tape turns are electrically insulated (`uncoupled') or in electrical contact (`coupled'). The AC losses computed with the $T$-$A$ formulation are compared to those computed with the $H$ formulation and the MEMEP method. In all cases, the superconductor is modeled as a material with power-law resistivity~\cite{Grilli:TAS14a}
.}}
\end{figure}
As mentioned in section~\ref{sec:T-A}, the $T$-$A$ formulation for thick conductors can be used for simulating not only individual conductors with rectangular cross section, but also stacks of coupled coated conductors. Here we present the validation of the model for the latter case, in particular for the stand-alone racetrack coil considered in~\cite{Pardo:TAS19,Benkel:arXiv19}. The coil is made of 4 cable turns, each made of 13 tape turns, which can be considered as electrically insulated or in electrical contact. For brevity, the two situations are referred to as `uncoupled' and `coupled', respectively. Fig.~\ref{fig:coils_alone_plot} shows the transport AC loss of such coils as a function of the normalized critical current. The transport current of each cable (each made of 13 turns) is \SI{2248.6}{\ampere} at \SI {500}{\hertz}. Different $I/I_{\rm c}$ ratios are obtained by changing $I_{\rm c}$. The figure presents a comparison between different models for the coupled and uncoupled case: the $T$-$A$ formulation (the uncoupled case is given by the original 1D model developed in~\cite{Zhang:SST17,Liang:JAP17}, whereas the coupled case is given by the approach presented in section ~\ref{sec:T-A}), the $H$ formulation and the MEMEP method. In the last two models, the tapes are simulated as individual objects, with different constraints on the current for the two coupling scenarios.

The results are in very good agreement with each other: with the exception of a few points at very low current ratios for the coupled case, the difference between the models is in the range of only a few~\%. Due to the non uniform current distribution among the tapes, the AC losses of the coupled case are about twice as high as those of the uncoupled case.
\begin{figure}[t!]
\centering
\includegraphics[height=7 cm]{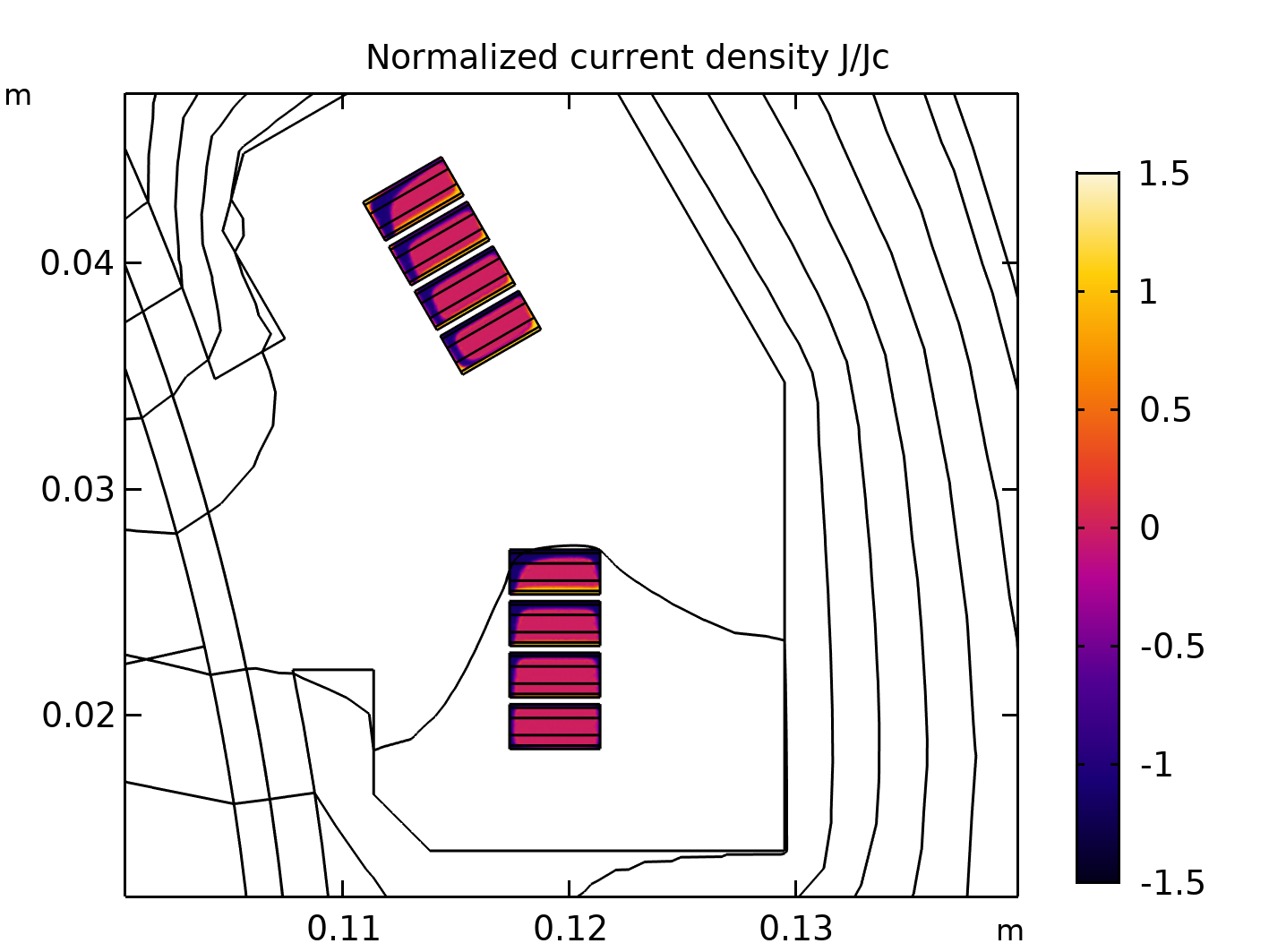}
\caption{\label{fig:J_distrib}{Current density distribution in two half-coils of the SUTOR motor's stator for the case of coupled turns.}}
\end{figure}

The $T$-$A$ formulation can be used to study the difference between the two coupling scenarios in electrical machines. As an example, Fig.~\ref{fig:J_distrib} shows the current density distribution in two half-coils of the stator of the superconducting motor considered in~\cite{Pardo:TAS19,Benkel:arXiv19}, whose design is based on that of the SUTOR motor~\cite{Oswald:PP12}. 
The average power loss dissipation of the stator coils in the coupled case are higher than in the uncoupled case, increasing from \SI{52.31}{\watt} to \SI{111.5}{\watt} at $\SI{65}{\kelvin}$, and from \SI{90.79}{\watt} to \SI{173.22}{\watt} at $\SI{77}{\kelvin}$.

\subsection{Quasi Critical State Model: One-step vs Full-cycle Simulations, Influence of Frequency}\label{sec:QCSM_results}
\begin{figure}[t!]
\centering
\includegraphics[width=6 cm]{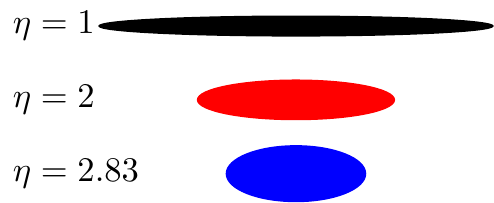}
\caption{\label{fig:ellipses_plot}Ellipses of different aspect ratio considered for the cross section of the superconductor. The black ellipse on the top has semi-axes $a=\SI{2}{\milli\meter}$ and $b=\SI{0.1}{\milli\meter}$. These values are, respectively, divided and multiplied by the factor $\eta$ to obtain the other two ellipses shown in the figure. The area of the ellipses is the same.}
\end{figure}

We compared the one-step and full-cycle simulations of the QCSM for the magnetization losses of a superconductor of elliptical cross section caused by an inclined magnetic field -- a case for which one-step simulations and Eq.~\eqref{eq:peak_losses} should not give the correct results for the reasons explained in Section~\ref{sec:QCSM}. We started with the simulation of an ellipse representative of the superconducting cross section of a Bi-2223 tape, with semi-axes $a=\SI{2}{\milli\meter}$ and $b=\SI{0.1}{\milli\meter}$ (black ellipse in Fig.~\ref{fig:ellipses_plot}), self-field critical current $I_{\rm c}=\SI{160}{\ampere}$ and constant $J_{\rm c}$. Then we considered two progressively narrower and thicker ellipses, while leaving  $I_{\rm c}$ and the total area unchanged: in particular, the semi-axes were divided and multiplied by the same factor $\eta$, which was set equal first to 2 (red ellipse in Fig.~\ref{fig:ellipses_plot}) and then to 2.83 (blue ellipse in Fig.~\ref{fig:ellipses_plot}).

\begin{figure}[t!]
\centering
\includegraphics[width=8 cm]{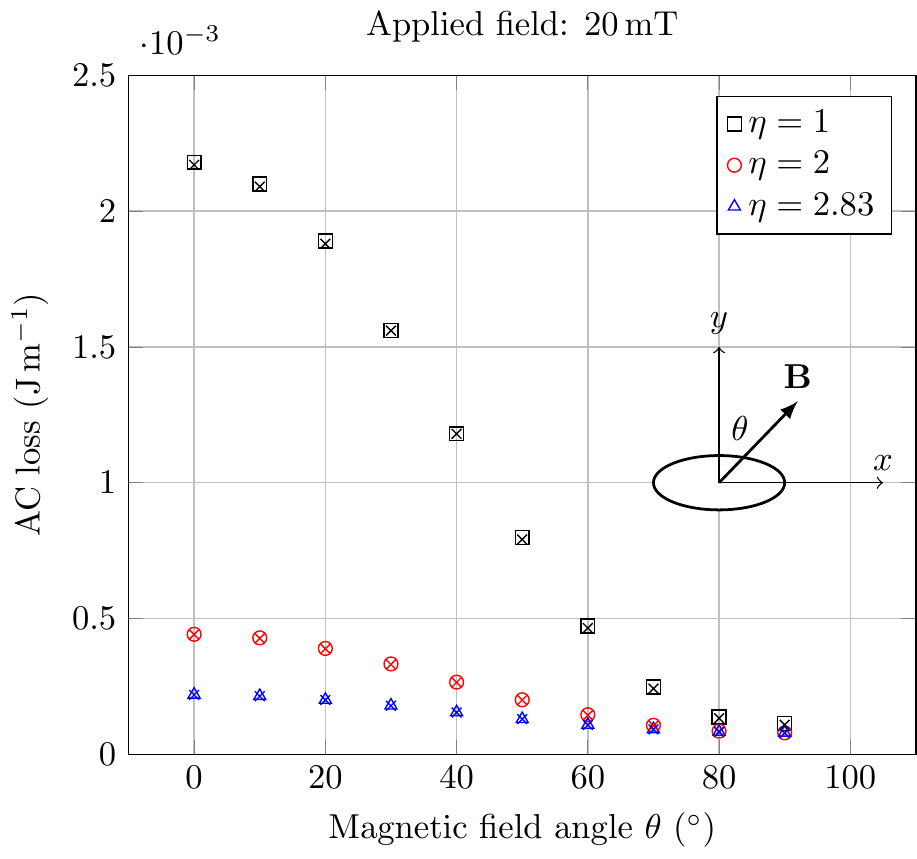}
\caption{\label{fig:inclined_20mT_plot}Magnetization AC losses in elliptical superconductors of different aspect ratio caused by an external AC field of $\SI{20}{\milli\tesla}$ at $\SI{50}{\hertz}$: comparison of simulation results obtained with the QCSM for different field orientations. Open symbols are the results of the simulation of the full AC cycle, the crosses are the result of the one-step calculation with equation~(\ref{eq:peak_losses}).}
\end{figure}

\begin{figure}[t!]
\centering
\includegraphics[width=8 cm]{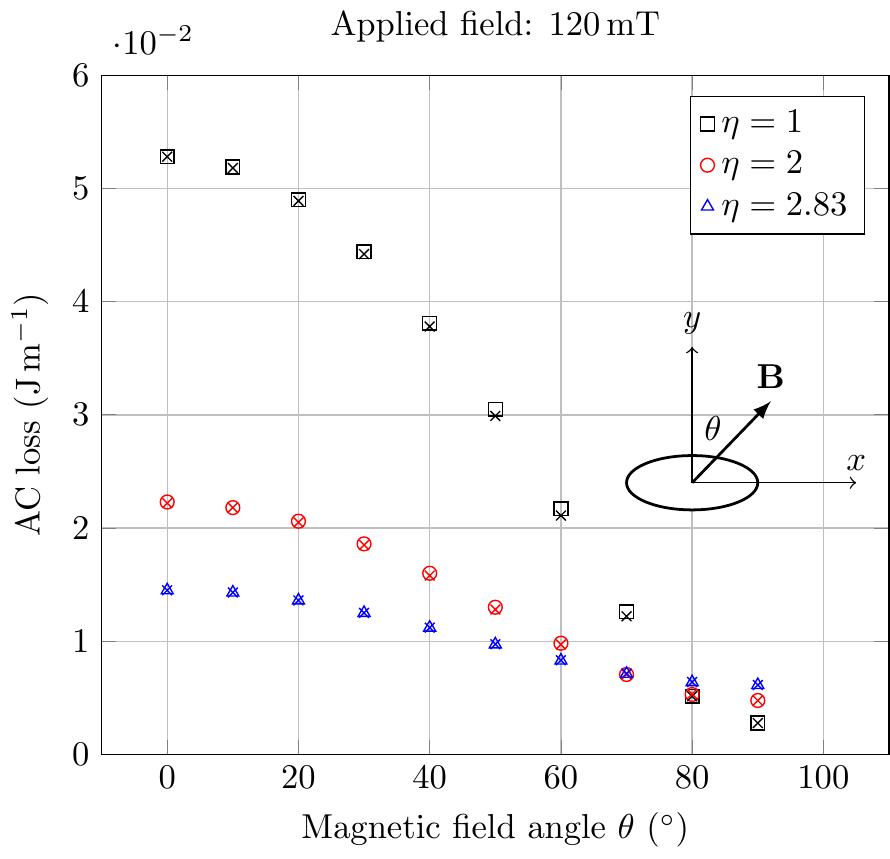}
\caption{\label{fig:inclined_120mT_plot}Magnetization AC losses in elliptical superconductors of different aspect ratio caused by an external AC field of $\SI{120}{\milli\tesla}$ at $\SI{50}{\hertz}$: comparison of simulation results obtained with the QCSM for different field orientations. Open symbols are the results of the simulation of the full AC cycle, the crosses are the result of the one-step calculation with equation~(\ref{eq:peak_losses}).}
\end{figure}

The AC applied field was set equal to 20 and $\SI{120}{\milli\tesla}$. These two values correspond to cases of partial and total penetration in the superconductor for $\theta=\SI{0}{\degree}$. In both cases, the orientation of the field was varied from 0 to $\SI{90}{\degree}$ (see insets of Figs.~\ref{fig:inclined_20mT_plot}-\ref{fig:inclined_120mT_plot} for the definition of the angle). The AC losses were computed as follows: for the full-cycle model, by integrating the product $\mathbf J \cdot \mathbf E$ over the superconductor's cross section and averaging over the second cycle; for the one-step model, by using Eq.~\eqref{eq:peak_losses}.
Figs.~\ref{fig:inclined_20mT_plot} and \ref{fig:inclined_120mT_plot} show that the losses calculated with the dynamic and static models are very similar, for both field amplitudes and all angles: the difference does not exceed $\SI{5}{\percent}$. 
These simulations seem to indicate that the differences in AC losses calculated with the full-cycle and one-step models are rather small, probably because the differences in current distribution are not great and only occur over a small part of the cycle; once the loss is averaged over a cycle, the effect is even smaller.
We also report that, due to similar reasons, the one-step calculation is also used in power-law based finite modeling for the fast calculation trapped magnetization of HTS bulks for sufficiently large values of $n$~\cite{Lousberg:SST09}.

\begin{figure}[t!]
\centering
\includegraphics[width=8 cm]{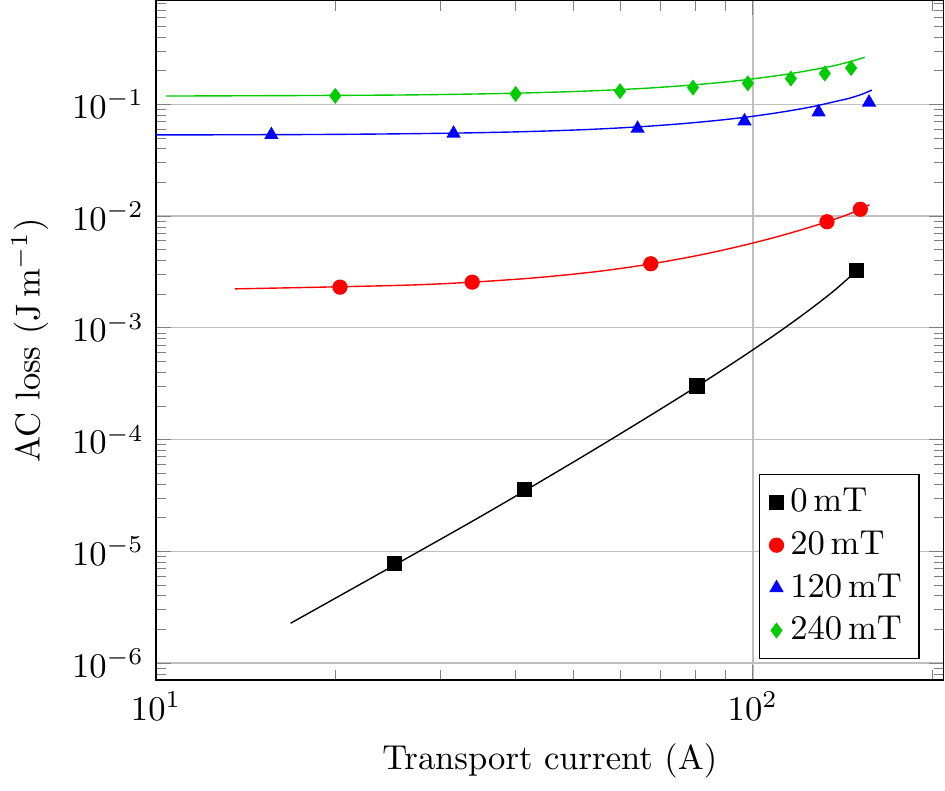}
\caption{\label{fig:AC_AC_FG_plot}AC losses of a thin ellipse ($\eta=1$, see Fig.~\ref{fig:ellipses_plot}) as a function of the AC transport current for different external AC magnetic fields. The field orientation is $\theta=\SI{0}{\degree}$ (see insets of Figs.~\ref{fig:inclined_20mT_plot}-\ref{fig:inclined_120mT_plot} for the definition of the angle). Full symbols and continuous lines represent the results obtained with the full-cycle and one-step simulations, respectively.}
\end{figure}

As a second comparison between the one-step and full-cycle simulations, we calculated the losses of a thin ellipse ($\eta=1$, see Fig.~\ref{fig:ellipses_plot}) under the simultaneous action of an AC transport current and an in-phase AC magnetic field with orientation $\theta=\SI{0}{\degree}$ (see insets of Figs.~\ref{fig:inclined_20mT_plot}-\ref{fig:inclined_120mT_plot} for the definition of the angle).
The results are given in Fig.~\ref{fig:AC_AC_FG_plot}, which shows that the predictions of the two models are in excellent agreement, except for the cases of high current and high field (top-right part of the figure). This is to be expected, because in those cases the sample is fully penetrated by the magnetic field resulting from the transport current and external magnetic field; in such situations, certain assumptions for the validity of eq.~(\ref{eq:peak_losses}) -- such as the existence of a kernel inside the sample where $A=0$ -- are no longer valid~\cite{Pardo:SST07}.

\begin{figure}[t!]
\centering
\includegraphics[width=8 cm]{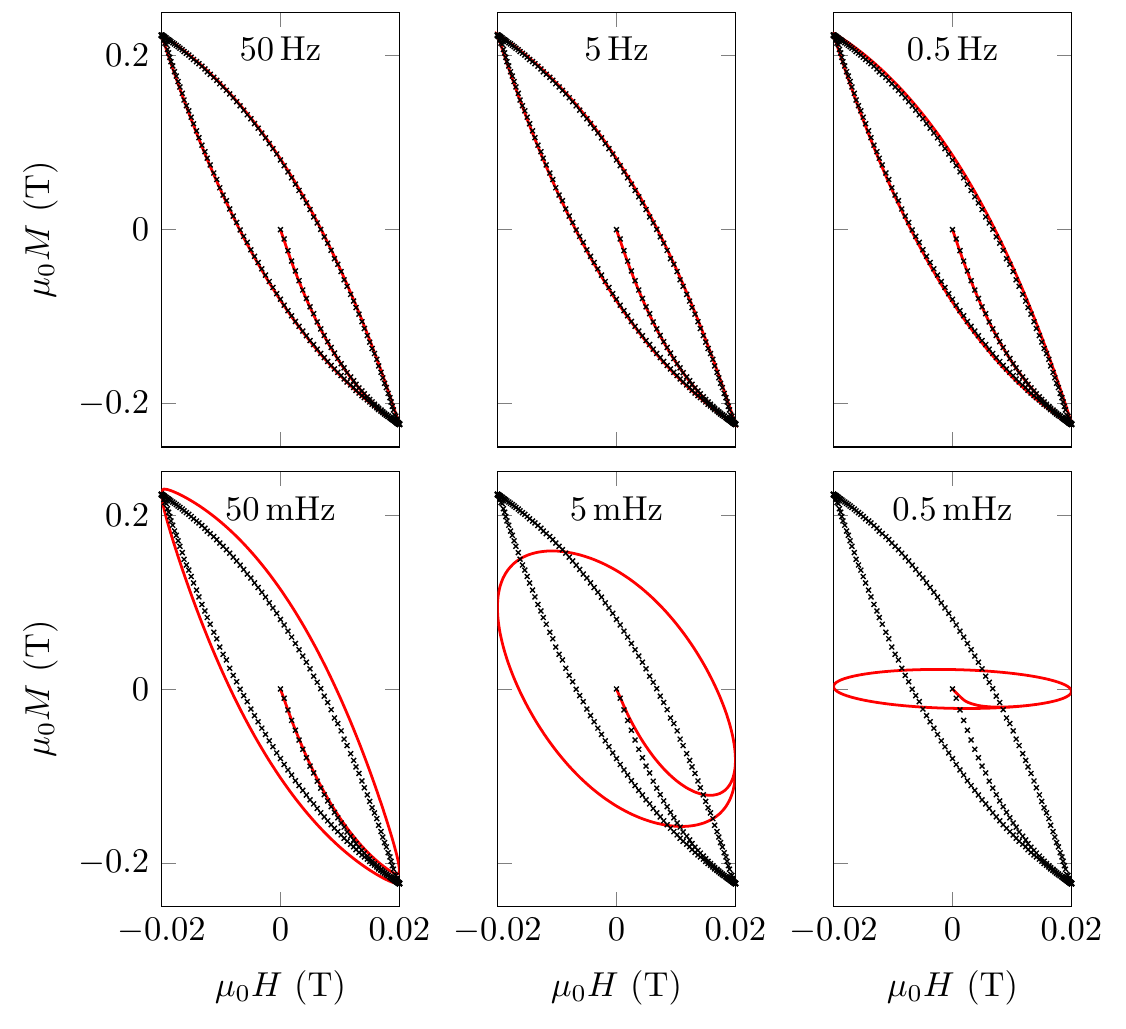}
\caption{\label{fig:MH_frequency_subplots3}{Magnetization cycles for the thin elliptical conductor of Fig.~\ref{fig:ellipses_plot}, subjected to a magnetic field of \SI{20}{\milli\tesla}, $\theta=\SI{0}{\degree}$}, of frequency varying between \SI{50}{\hertz} and $\SI{0.5}{\hertz}$. Results obtained with the QCSM (red lines) and with the `pure' CSM (black crosses). The latter is independent of frequency.}
\end{figure}

Finally, with the full-cycle model, we varied the frequency of the external AC field (\SI{20}{\milli\tesla}, $\theta=\SI{0}{\degree}$) applied to the thin elliptical conductor of Fig.~\ref{fig:ellipses_plot}.  As shown by the red lines in Fig.~\ref{fig:MH_frequency_subplots3}, if the frequency is sufficiently low, the results depart from those obtained with a pure critical state model (black crosses) and the calculated losses strongly depend on the frequency. 
This is due to the fact that rate-independent results are obtained only when the generated electric field is sufficiently large (compared to the chosen $E_0$ in~\eqref{eq:EJ}). This is not the case when the frequency of the applied magnetic field is too low: in that case, the results depart from that of a `pure' critical state model. 

Because of this dependence on the rate of the excitation, we called this model a {\it quasi} critical state model. Care should therefore be taken, if the model is used to simulate situations with slowly changing fields and/or transport currents.

\section{Conclusion}

With this contribution, we have  extended the $T$-$A$ formulation to the case of thick superconductors. This extension also allows simulating stacks (or windings) of electromagnetically coupled HTS coated conductors. The formulation has been applied to calculate the losses of the stator coils of a superconducting motor and to compare the case of uncoupled and coupled turns.

Full-cycle and one-step simulations of the quasi critical state model have been compared in terms of AC loss calculation. In the case of individual superconducting tapes subjected to external AC magnetic fields, the difference of the AC loss results is rather marginal. In that case, the static model can be used to rapidly evaluate the AC losses of superconducting tapes. In the case of simultaneous AC current and AC field, however, the losses calculated with the one-step model can be significantly different from those calculated with full-cycle simulations, particularly in the case of high current and high fields. In the low-current and low-field range, however, the agreement between the two types of simulation is excellent and one-step simulations can be used for a very rapid estimation of the losses.

In addition, the full-cycle simulations with the quasi critical state model revealed a dependence of the AC loss results on the frequency of the AC excitation, if the frequency range is sufficiently large. This dependence should be taken into account, especially for the simulation of slow varying fields.

\section*{Appendix}

Imposing a transport current $I(t)$ by using~\eqref{eq:current} in a 2D superconductor of arbitrary shape  proved to be not straightforward in COMSOL Multiphysics. In particular, the simple imposition of a global constrain proved to be ineffective.
A procedure that works is the following.
First, define a new variable $Y$ in a newly created {\it Global ODE} physics:
\begin{equation}
\verb!Y=intop1(J)-I(t)!
\end{equation}
where \verb intop1 ~is an integral operator and \verb intop1(J) ~represents the calculated integral of the current density on the superconductor's cross section.
Then, in the {PDE} physics for the $\mathbf T$ potential, add a {\it Weak Contribution} on the superconductors boundary
\begin{equation}
\verb!(Y-Ttan)*test(Ttan)!
\end{equation}
where \verb Ttan=Tx*tx+Ty*ty ~is the projection of the two components $(T_x,T_y)$ of the potential $\mathbf T$ on the superconductor's boundary.

\begin{figure}[t!]
\centering
\includegraphics[width=8 cm]{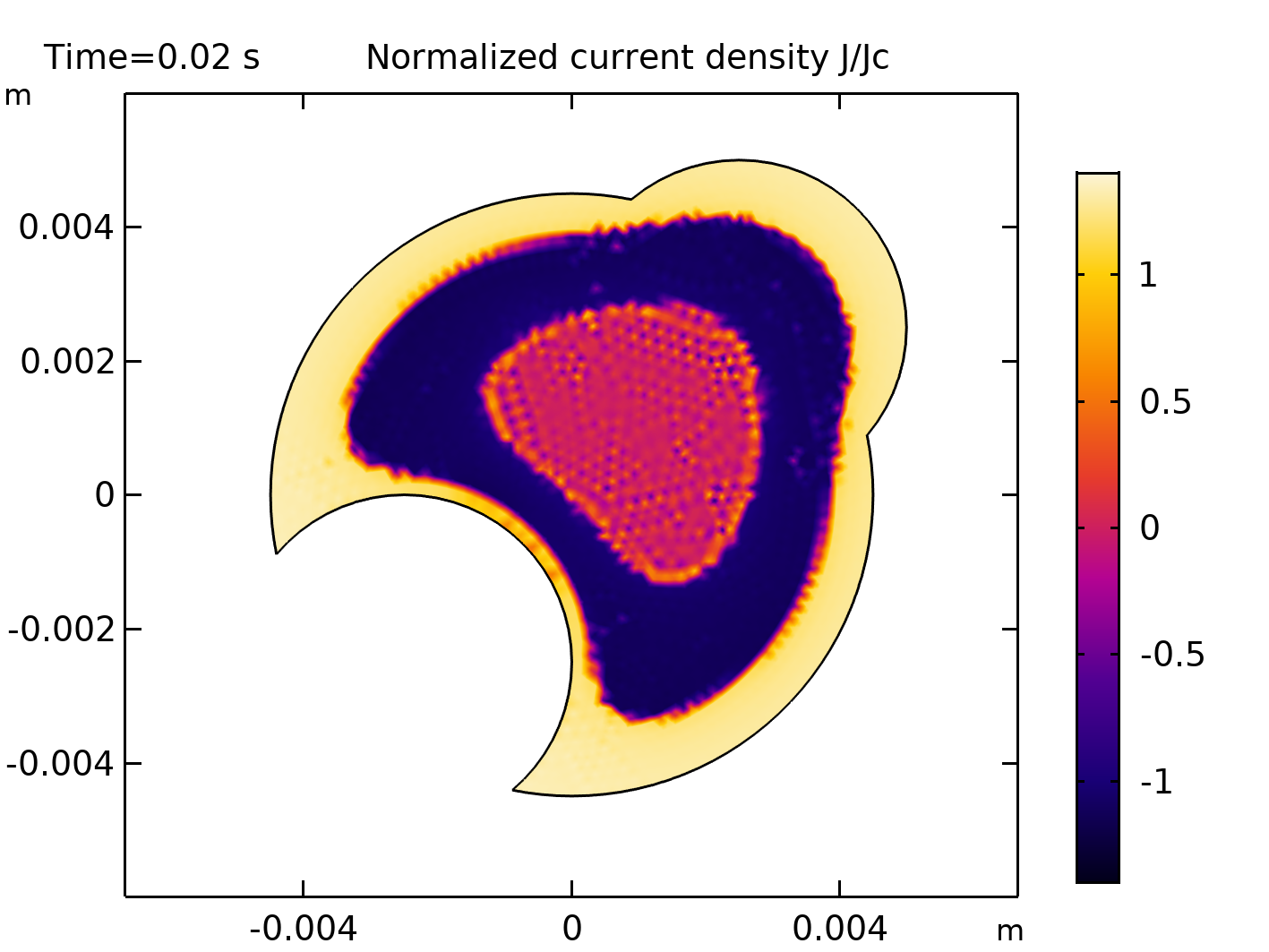}
\includegraphics[width=8 cm]{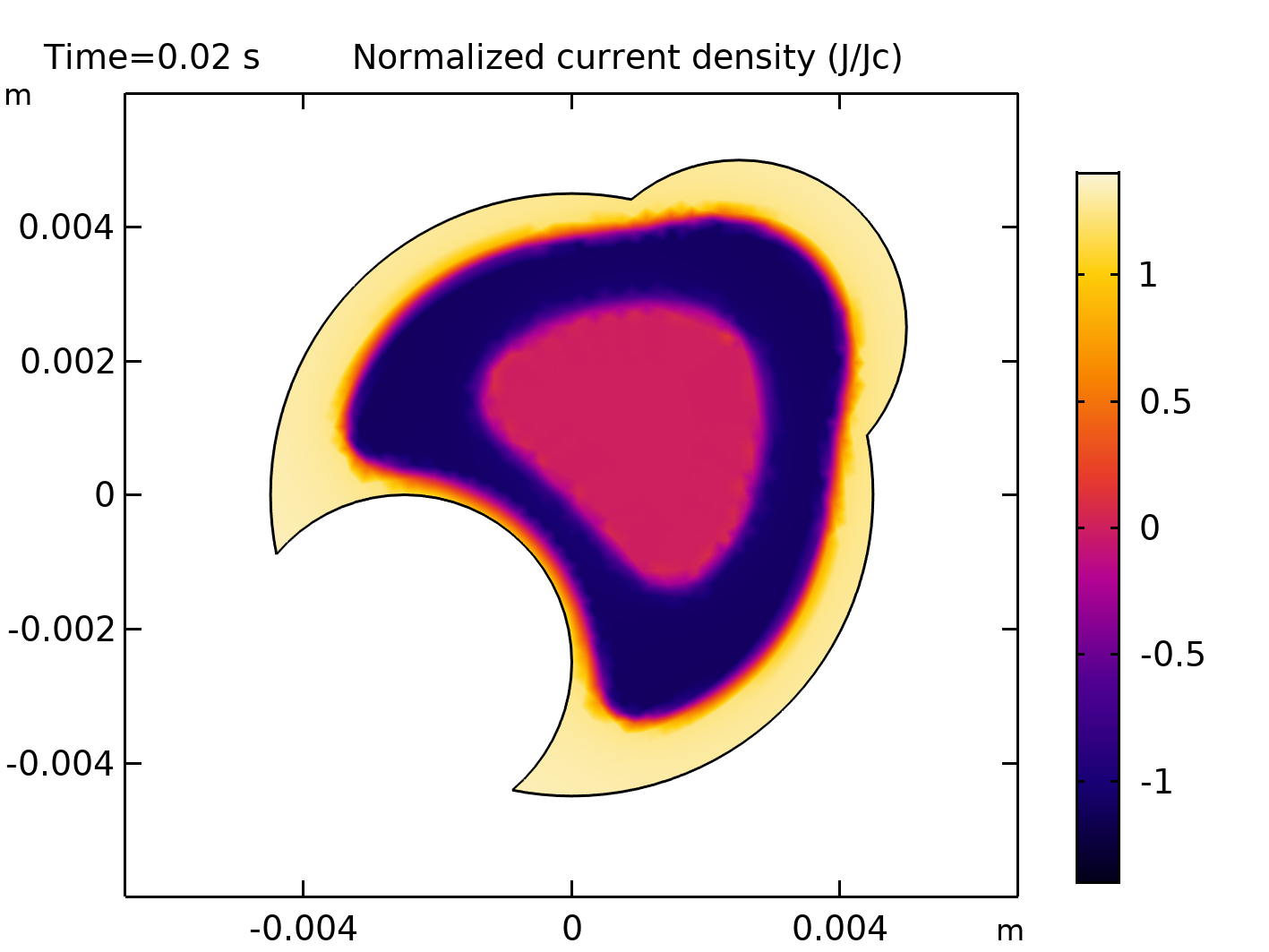}
\caption{\label{fig:arbitrary}Comparison between $T$-$A$ and $H$ formulations: current density distribution in a superconductor of non-trivial cross section carrying an AC transport current. The distributions are taken at the end of the AC cycle.}
\end{figure}

This procedure was successfully tested on several shapes, and an example is reported in Fig.~\ref{fig:arbitrary}. The figure shows the distribution of the normalized current density in a superconductor of arbitrary shape at the end of an AC cycle of  transport current. The results obtained with the $T-A$ formulation (top) are compared with those obtained with the $H$ formulation. The agreement is very good and the calculated AC losses (not shown here) are identical.

%


\ifCLASSOPTIONcaptionsoff
  \newpage
\fi





\end{document}